\documentclass[a4paper,fleqn,review]{cas-dc}
\usepackage[numbers]{natbib}
\usepackage{color}

\usepackage{array}
\usepackage{booktabs}
\usepackage{caption}

\usepackage{wrapfig} 

\usepackage{listings}

\usepackage{cite} 
\usepackage{float}
\usepackage{multirow}
\usepackage{booktabs}
\usepackage{comment}
\usepackage{rotating}
\usepackage{xspace}
\usepackage{tcolorbox}
\tcbuselibrary{skins}
\tcbuselibrary{listings}

\usepackage{makecell}

\usepackage{tikz}
\usepackage{wasysym}
\usepackage{fontawesome}
\usepackage{tabularx}
\usepackage{graphicx}
\usepackage{xcolor}
\usepackage{array}

\usepackage{threeparttable}

\usepackage{arydshln} 

\definecolor{SeaGreen}{RGB}{46, 139, 87}
\definecolor{CornflowerBlue}{RGB}{100, 149, 237}
\definecolor{RoyalPurple}{RGB}{120, 81, 169}
\definecolor{Aquamarine}{RGB}{127, 255, 212}

\def\tsc#1{\csdef{#1}{\textsc{\lowercase{#1}}\xspace}}
\tsc{WGM}
\tsc{QE}

\begin{document}
\let\WriteBookmarks\relax
\def\floatpagepagefraction{1}
\def\textpagefraction{.001}
\let\printorcid\relax 

\shorttitle{~}    

\shortauthors{Suo et al.}

\title[mode = title]{Assessing the Capability of Android Dynamic Analysis Tools to Combat Anti-Runtime Analysis Techniques}  

\author[1,3]{Dewen Suo}
\author[2]{Lei Xue}
\author[2]{Weihao Huang}
\author[2]{Runze Tan}
\author[1]{Guozi~Sun}
\cormark[1]

\address[1]{Nanjing University of Posts and Telecommunications, No.9, Wenyuan Road, Yadong New District, Nanjing, 210023, China} 
\address[2]{Sun Yat-sen University, No. 66, Gongchang Road, Guangming District, Shenzhen, Guangdong 518107, China} 
\address[3]{China Telecom Shandong Branch, Jinan, 250000, China.} 
\cortext[1]{Corresponding author: Guozi~Sun (Email: sun@njupt.edu.cn).}  

\begin{abstract}
As the dominant mobile operating system, Android continues to attract a substantial influx of new applications each year. However, this growth is accompanied by increased attention from malicious actors, resulting in a significant rise in security threats to the Android ecosystem. Among these threats, the adoption of Anti-Runtime Analysis (ARA) techniques by malicious applications poses a serious challenge, as it hinders security professionals from effectively analyzing malicious behaviors using dynamic analysis tools. ARA technologies are designed to prevent the dynamic examination of applications, thus complicating efforts to ensure platform security. This paper presents a comprehensive empirical study that assesses the ability of widely-used Android dynamic analysis tools to bypass various ARA techniques. Our findings reveal a critical gap in the effectiveness of existing dynamic analysis tools to counter ARA mechanisms, highlighting an urgent need for more robust solutions. This work provides valuable insights into the limitations of existing tools and highlights the need for improved methods to counteract ARA technologies, thus advancing the field of software security and dynamic analysis.
\vspace{-1em}
\end{abstract}




\maketitle


\section{Introduction}
\label{sec:intro}

Android is the leading mobile operating system, commanding an 80\% market share in the smartphone sector~\citep{idc-smartphone}. However, this dominance also makes it a prime target for malicious attacks, which pose significant risks to user privacy, data security, and overall system integrity~\citep{Shen2021ALT,Yang2024FromGT}. As a result, researchers and security analysts have developed a variety of sophisticated tools and techniques aimed at Android malware detection and mitigation.

Malicious applications can infringe upon user privacy, steal sensitive information, and result in economic losses. In response to these threats, security researchers and analysts have been actively studying Android malware and developing effective countermeasures. Meanwhile, to conceal malicious payloads, malware developers have increasingly turned to anti-analysis techniques that obfuscate or hinder analysis processes~\citep{Sihag2021ASO}. Consequently, the academic community has made significant efforts to develop powerful tools for in-depth application analysis.

Traditional anti-analysis techniques typically focus on hindering static analysis by increasing its complexity. One of the most widely used methods is code obfuscation~\citep{Chan2004AdvancedOT,liang1998dynamic,Huang2014TypeBasedTA}. Code obfuscation increases the complexity of the code by modifying its structure and variable names, making static code analysis more challenging.

In contrast, Anti-Runtime Analysis (ARA) technology is a relatively new category of anti-analysis techniques that specifically target dynamic analysis rather than static methods~\citep{Berlato2020ALS,Lita2018AntiemulationTI}.  Initially, ARA technology was designed for legitimate purposes, such as protecting intellectual property and preventing reverse engineering~\citep{Jang2019RethinkingAT,Haupert2018HoneyIS}.  However, the adoption of ARA technology by malicious software developers has complicated the work of security analysts~\citep{Tam2017TheEO,Lita2018AntiemulationTI}.  By employing ARA techniques, these developers can obscure the detection of malicious behaviors, effectively prolonging the lifecycle of their malicious applications.  In order to effectively analyze the behavior of malicious applications, Android dynamic analysis tools must possess the capability to handle ARA technology.

Previous studies have systematically evaluated the performance of Android static analysis tools, particularly their ability to handle code obfuscation~\citep{Hammad2018ALE,Nellaivadivelu2020BlackBA,Soi2023CanYS,Nawaz2022OnTE}. These studies have confirmed that code obfuscation can significantly impact the effectiveness of static analysis tools~\citep{Gao2024ACS}. However, there is a notable gap in research concerning the performance of Android dynamic analysis tools and their ability to handle ARA techniques.

To address this gap, we evaluated six well-known Android dynamic analysis tools using a dataset of 993 benign and 991 malicious applications.  Our primary objective was to assess the effectiveness of existing Android dynamic analysis tools in handling ARA techniques, using code coverage as the evaluation metric. Specifically, we aimed to evaluate the effectiveness of dynamic analysis tools in countering ARA techniques through multiple controlled experiments by comparing the code coverage of applications in their original form with that achieved when analyzed using dynamic analysis tools. This study aims to assist security professionals in better understanding the limitations and capabilities of current dynamic analysis tools, enabling them to make more informed decisions when selecting tools. Additionally, we seek to highlight the gaps in the existing tools to promote technological advancements in the field.

To highlight the practical implications of our study, we summarize three representative findings below:  
(1) None of the evaluated dynamic analysis tools were able to effectively handle ARA techniques, regardless of whether the target applications were benign or malicious;
(2) Increasing the number of deployed ARA techniques generally led to stronger resistance against analysis. Notably, the most significant drop in code coverage occurred between applications with zero ARA techniques and those with one to five techniques. Although deploying more than five techniques continued to enhance resistance, the marginal effect diminished;  
(3) When considering both analysis capability and runtime efficiency, DroidDissector~\citep{Muzaffar2023DroidDissectorAS} demonstrated the best overall performance among the six evaluated tools. Further detailed results are discussed in Section~\ref{sec:DATA_ANALYSIS_AND_RESULTS}, encompassing a broader spectrum of technical findings.

To guide our research, we have distilled three key research questions, including:

\textit{\underline{Q1}: Can dynamic analysis tools effectively handle ARA technology?} 

\textit{\underline{Q2}: What are the differences in the impact of various categories and quantities of ARA technology on dynamic analysis tools?} 

\textit{\underline{Q3}: How efficient are dynamic analysis tools when processing a large-scale dataset of applications protected by ARA techniques?}

We will provide a detailed discussion of the three questions we have posed in section \ref{sec:DATA_ANALYSIS_AND_RESULTS}.

However, obtaining convincing answers to these questions is not straightforward, as it requires a large-scale investigation and the resolution of key technical challenges and practical constraints.

\textit{C: Lack of an automated evaluation framework for ARA detection and dynamic analysis tools assessment.}
Evaluating dynamic analysis tools for their ability to handle ARA techniques is time-consuming. Currently, there is a lack of an automated framework that can effectively detect ARA techniques in APKs and assess how well the tools handle them. Such a framework needs to be highly scalable to facilitate future evaluations of additional tools.

In addition to this core challenge, our study must also contend with practical constraints, such as the difficulty of measuring code coverage without access to the source code of the target APKs. Since Android applications distributed via platforms like Google Play are typically available only in compiled form, analysis must be performed in a black-box setting. To address this, we integrated ACVTool~\citep{pilgun2020acvtool} to enable runtime code coverage measurement.

Section~\ref{sec:Research_Methodology} explains how this study addresses the above challenge and mitigates these constraints. This paper’s key contributions are as follows:

$\bullet$~\textbf{Bridging a Research Gap}: This study is the first to systematically evaluate the ability of dynamic analysis tools to handle ARA techniques, thus filling a critical gap in the existing literature.

$\bullet$~\textbf{Large-Scale Dataset}: We have created a dataset consisting of 993 benign and 991 malicious applications, and generated corresponding ARA analysis reports for each. Our dataset, available at https://github.com/dfpp/Anti-ARA, provides a valuable resource for future research.

$\bullet$~\textbf{Practical Implications}: By identifying the strengths and weaknesses of current Android dynamic analysis tools in dealing with ARA techniques, this research offers essential guidance to security professionals, IT auditors, and mobile application developers in strengthening the security posture of Android systems.

\textit{Roadmap}: The remaining parts of this paper are organized as follows: Section \ref{sec:background} provides background information on Android dynamic analysis and ARA technology.  Section \ref{sec:Research_Questions} discusses in detail the research questions addressed in this study.  Section \ref{sec:Research_Methodology} introduces the research methodology.  Results and findings are reported in Section \ref{sec:DATA_ANALYSIS_AND_RESULTS}.  Section \ref{sec:DISCUSSION} discusses the results and provides recommendations for enhancing dynamic analysis tools.  Threats to validity are presented in Section \ref{sec:THREATS_TO_VALIDITY}.  Finally, this paper outlines related work (Section \ref{sec:RelatedWork}) and draws conclusions (Section \ref{sec:CONCLUSION}).

\vspace{-1em}
\section{Background}
\label{sec:background}

This section provides a brief overview of Android dynamic analysis and Anti-Runtime Analysis (ARA) technologies, offering readers essential context to better understand the subsequent discussion in the paper.
\vspace{-1em}
\subsection{Android Dynamic Analysis}

Android dynamic analysis refers to a method of analyzing the security and performance characteristics of an application by observing and recording its behavior and interactions during runtime.  Compared to static analysis, dynamic analysis has the advantage of capturing the application's actual runtime behavior, which enables a more comprehensive assessment of its security and performance~\citep{Wu2019AnalysesFS}.

The widespread use of obfuscation techniques has significantly reduced the effectiveness of static analysis for app scrutiny~\citep{Xie2023PreciseAE,Tang2021AndroidMO,Conti2022ObfuscationDI,Guo2022ASO}.  Obfuscation techniques modify the static code and resources of an application, making them harder to understand and analyze. As a result, static analysis becomes less efficient and accurate~\citep{Hammad2018ALE}.  In contrast, dynamic analysis enables the real-time observation and recording of an application's behavior at runtime, unaffected by obfuscation techniques.  Thus, dynamic analysis can better address the challenges posed by obfuscation techniques, enhancing both the comprehensiveness and accuracy of the analysis~\citep{ruggia:hal-04378941}.  Consequently, the importance of dynamic analysis in the context of mobile application security is becoming increasingly evident.

Dynamic analysis tools can vary in their workflows, but a common approach involves preprocessing the APK file before performing dynamic analysis. Preprocessing allows the tool to capture runtime information, facilitating more effective analysis. During the dynamic analysis process,  researchers aim to trigger as many app functionalities as possible to achieve a comprehensive evaluation.  Researchers can utilize automated tools such as Monkey~\citep{Monkey}, Delm~\citep{Hu2024EnhancingGE}, or Fastbot~\citep{LvPZSLY22} to efficiently explore the app in an automated manner. In this paper, we focus on evaluating this type of dynamic analysis tool.

\vspace{-0.5em}
\subsection{Anti-Runtime Analysis Techniques}

ARA techniques refer to a set of methods used in Android applications to protect them from dynamic runtime analysis attacks.  These attacks involve analyzing the behavior of the application during execution to identify vulnerabilities or extract sensitive information.

The emergence of these ARA techniques is driven by the differing motivations of benign and malicious application developers.   Benign developers often use ARA technology to safeguard their intellectual property, prevent reverse engineering, and protect against unauthorized tampering.   By employing ARA technology, they can protect their code from unauthorized analysis or modification.

Malicious developers, in contrast, exploit ARA technology to hide their malicious activities, evade detection and countermeasures, and enhance the persistence of their applications.   The use of ARA technology hinders security researchers and analysts from understanding application behavior and detecting malicious activities.

\begin{figure}[t]
\centerline{\includegraphics[width=1.0\linewidth]{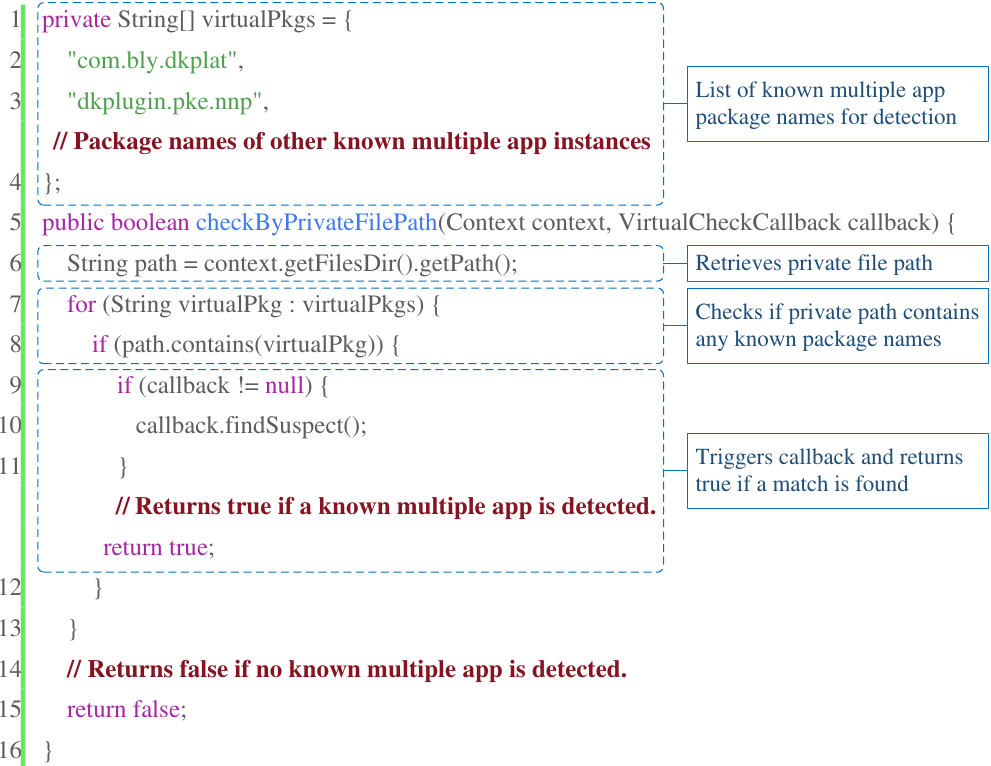}}
\vspace{-0.5em}
\caption{A Motivating Example}
\label{fig:CODE}
\vspace{-1.5em}
\end{figure}

In this paper, we consider five types of ARA techniques:

$\bullet$~\textbf{Anti-Debugging (AD)}: Anti-Debugging is a technique used to protect applications from reverse engineering attacks~\citep{Apostolopoulos2021ResurrectingAA}.  This involves inserting anti-debugging code into the application, making it difficult for analysts to use debuggers or other tools to debug and reverse-engineer the application.  Properly deploying anti-debugging techniques can significantly enhance the security of the application, reinforcing its resilience against malicious attacks~\citep{Wan2018ADA}.

$\bullet$~\textbf{Anti-Hooking (AH)}: Anti-Hooking refers to techniques employed to prevent attackers from using hooking technologies to monitor, modify, or tamper with the behavior of an application, which could facilitate malicious attacks or the theft of sensitive information~\citep{Haupert2018HoneyIS}.  Commonly used tools include Xposed~\citep{XposedRepo} and Frida~\citep{FridaRepo}.


$\bullet$~\textbf{Anti-Tampering (AT)}: Anti-Tampering prevents malicious attackers from modifying or repackaging applications. Implementing such mechanisms protects apps from unauthorized changes, ensuring security and integrity~\citep{Berlato2020ALS,Li2018RebootingRO,Ruggia2021RepackMI}.

$\bullet$~\textbf{Root Detection (RD)}: In Android, "Rooting" is the process of obtaining superuser privileges. With Root access, users bypass system restrictions, modify settings, delete pre-installed apps, and install unverified ones~\citep{NguyenVu2017AndroidRA}. Root detection helps apps identify rooted devices to take actions safeguarding security.

$\bullet$~\textbf{Virtual Environment Detection (VED)}: This technique aims to detect if the application is running within a virtual environment ~\citep{GuerraManzanares2019DifferencesIA,Sahin2018ProteusDA}.  Malicious applications may use this information to decide whether to alter their behavior to evade detection~\citep{Wang2017DroidAntiRMTC,Afonso2018LumusDU}, while benign applications may check for a virtual environment to confirm that the application is running in a secure environment~\citep{Jang2019RethinkingAT,Xu2021AndroidOP,Hong2017DefiningAD}.

To facilitate a better understanding of ARA techniques, we present a code snippet in Figure \ref{fig:CODE} as a motivating example, taken from a real-world application.

In this code fragment, the goal of the application is to check whether the current runtime environment is an application-level virtualization environment. Specifically, the code achieves this by examining file paths. The basic idea is that the host application redirects the client application’s APK to its own directory when loading it. In line 6, the application uses the getPath function to retrieve the path to the application's internal storage directory and checks whether the path contains common package names associated with multiple applications (lines 7-8). If a common package name is detected, the application invokes a callback function to perform the corresponding action (lines 9-11). This detection method requires a pre-prepared list of known application package names (lines 1-4).

ARA techniques are complex technologies with various implementation methods. Here, due to space constraints, we provide only a concise description of ARA technology. For a more detailed explanation and technical implementation of ARA technology, we refer readers to our paper \citep{Suo2024ARAPDA}, which offers an in-depth analysis. Additionally, the corresponding open-source project on GitHub can further aid in understanding the practical aspects of ARA technology. By combining the information from the paper and the project, readers will gain a comprehensive understanding of ARA techniques.

\vspace{-1em}
\section{Research Questions}
\label{sec:Research_Questions}
{

This study conducts a large-scale empirical evaluation of dynamic analysis tools in the context of ARA techniques used in Android applications. Our investigation is driven by three research questions, each targeting a different aspect of the problem. These questions not only shape our experimental design but also serve as a practical guide for tool developers, application authors, and security analysts in selecting or improving dynamic analysis strategies.

\vspace{-0.3em}
\begin{tcolorbox}[left=2mm, right=2mm, top=1mm, bottom=1mm]
\faLightbulbO \ \  {\textbf{Q1: Can dynamic analysis tools effectively handle ARA technology?}}
\end{tcolorbox}
\vspace{-0.3em}

\textit{Reasons.}
In recent years, ARA techniques have been widely adopted by both benign and malicious developers to resist reverse engineering and hinder dynamic inspection. As security analysts increasingly rely on dynamic analysis tools to understand application behavior and uncover vulnerabilities, it becomes essential to assess whether these tools are capable of overcoming such protections.

\textit{Objective.}
Q1 evaluates the ability of various dynamic analysis tools to mitigate or bypass ARA techniques during analysis.

\textit{Method.}
To ensure general applicability, especially in the context of closed-source tools, we use code coverage as the primary evaluation metric. Specifically, we compare coverage between the original APKs (unprocessed) and those instrumented by dynamic analysis tools. If a tool effectively handles ARA logic, it should be able to increase coverage relative to the original baseline.

\textit{Implication.}
Q1 results guide dynamic analysis tool developers to identify current limitations and improve robustness. For security analysts, this information aids selecting tools that deliver reliable results in real-world protected application settings.

}

\vspace{-0.3em}
\begin{tcolorbox}[left=2mm, right=2mm, top=1mm, bottom=1mm]
\faLightbulbO \ \  {\textbf{Q2: What are the differences in the impact of various categories and quantities of ARA technology on dynamic analysis tools?}}
\end{tcolorbox}
\vspace{-0.3em}

\textit{Reasons.}
Dynamic analysis tools can be affected differently by various categories of ARA techniques, owing to differences in design complexity, implementation maturity, and runtime behavior. Some techniques (e.g., virtualization-related checks) are lightweight and widely adopted, while others (e.g., anti-hooking mechanisms) may be more complex or less reliable in real-world deployment. It is therefore necessary to analyze how categories and \emph{quantities} of ARA techniques influence a tool’s ability to explore application behavior.

\textit{Objective.}
Q2 seeks to assess (i) how different \emph{categories} of ARA techniques affect dynamic analysis tools, and (ii) how the \emph{quantity} of deployed ARA techniques within an APK impacts tool effectiveness.

\textit{Method.}
We categorize applications by the \emph{type} and \emph{number} of ARA methods they deploy, and then analyze the resulting changes in \emph{code coverage} under dynamic analysis. In theory, more protections may enhance resistance but also introduce runtime overhead or diminishing returns; we empirically examine these trade-offs by comparing coverage distributions across categories and increasing counts of deployed methods.

\textit{Implication.}
The results reveal the relative impact of specific ARA categories against current tools and the effect of increasing protection quantity. For tool developers, identifying which categories (e.g., VED) are most obstructive helps prioritize mitigation strategies (e.g., more faithful environment emulation). For security analysts, unexpectedly low coverage can serve as a diagnostic signal of high-impact techniques, suggesting the need for alternative tools or manual intervention. For application developers, the findings indicate which techniques most effectively hinder analysis, informing more robust protection designs.

\vspace{-0.3em}
\begin{tcolorbox}[left=2mm, right=2mm, top=1mm, bottom=1mm]
\faLightbulbO \ \  {\textbf{Q3: How efficient are dynamic analysis tools when processing a large-scale dataset of applications protected by ARA techniques?}}
\end{tcolorbox}
\vspace{-0.3em}

\textit{Reasons.}
While resilience to ARA techniques is critical, \emph{efficiency} is equally important—especially when analyzing large-scale datasets comprising hundreds or thousands of APKs. A tool that can bypass protections but fails to scale efficiently becomes impractical in real-world workflows.

\textit{Objective.}
Assess the \emph{efficiency and reliability} of representative dynamic analysis tools on a large-scale dataset of ARA-protected apps, providing actionable guidance for time- and resource-constrained analysis settings.

\textit{Method.}
We evaluate efficiency along two key axes. (i) \emph{Preprocessing (instrumentation) time}: the time required to modify each APK prior to execution, which can become a bottleneck at scale. (ii) \emph{Validity and success rates}: whether instrumentation yields a valid APK (installable and executable on the analysis platform), and the subsequent installation/launch success during analysis. For each app–configuration, we perform three independent runs and report averages to reduce variance.

\textit{Implication.}
The results help security analysts select tools that remain practical under time constraints and operational realities, and they guide tool developers to optimize preprocessing steps and improve stability for large-scale use. Together with RQ1 and RQ2, RQ3 complements effectiveness-oriented findings with practical considerations for real-world deployment.

Together, these three questions provide a comprehensive framework for understanding the current capabilities and limitations of dynamic analysis tools in ARA-aware environments. They also lay the foundation for future improvements in tool design, application protection strategies, and analysis methodology.

\vspace{-1em}
\section{Study Methodology}
\label{sec:Research_Methodology}

This section outlines the methodology adopted for evaluating the effectiveness of dynamic analysis tools in handling Anti-Runtime Analysis (ARA) technologies. The core idea behind our research is that dynamic analysis tools, by preprocessing Android APKs, enhance the analyzability of the applications and increase code coverage during analysis. This is achieved by inserting monitoring and logging code into the APK, which alters the application’s original behavior, preventing it from detecting the debugging environment or other runtime anomalies. As a result, the application is unable to execute self-defense mechanisms, such as terminating itself or disabling certain features, which would otherwise hinder the analysis process.

Our approach begins with the selection of a dataset of benign and malicious applications, followed by ARA technology detection using ARAP. Next, we use ACVTool for code coverage measurement and conduct experiments using a modular evaluation framework to assess the capabilities of various dynamic analysis tools. By comparing the performance of these tools under controlled conditions, we aim to provide valuable insights into their effectiveness in counteracting ARA techniques.

\subsection{Study Subjects}
We first collected 1,000 benign applications and 1,000 malicious applications for constructing the dataset. To ensure the reliability of the data, we obtained samples from AndroZoo~\citep{Allix2016AndroZooCM} and used the "vt\_detection" values as the criterion to classify the samples as benign or malicious. AndroZoo is a renowned large-scale repository for collecting and storing Android application samples, and it includes a "vt\_detection" value indicating how many antivirus programs flagged the app as malicious (not a percentage). In this study, we considered an app with a "vt\_detection" value of 0 as benign, while an app with a "vt\_detection" value of 10 or more, indicating detection by over ten antivirus programs, was considered malicious.

To ensure that our dataset reflects the most current real-world scenarios and to optimize the performance of the dynamic analysis tools, we focused only on applications from AndroZoo that were uploaded after 2020. This decision was made to capture more recent trends in app development and malicious behavior. Furthermore, in this study, we aimed to construct a dataset as large as possible to reduce the impact of randomness on our results. However, due to the high computational and time overhead of dynamic analysis tasks, the dataset size had to be balanced against practical constraints. Specifically, we conducted 21 dynamic analysis runs for each APK sample—covering six different analysis tools (see Section 4.4 for details) plus one baseline, with three repeated experiments for each (i.e., 6+1=7 configurations × 3 runs = 21 runs per APK). Given this substantial testing cost, it was necessary to limit the overall dataset size. After careful consideration, we settled on a dataset that balances coverage and feasibility.

Additionally, to prevent potential contamination from outdated "vt\_detection" values in AndroZoo, we re-scanned all selected APKs using VirusTotal. This re-scanning process led to slight adjustments in the dataset, resulting in 993 benign applications and 991 malicious applications. While we initially aimed for 1,000 applications in each category, this adjustment was necessary to ensure the accuracy and reliability of the data. Given the large-scale nature of the dataset and the rigorous testing procedures employed, our dataset is among the largest used in dynamic analysis tool evaluations, especially in comparison to other similar studies. Ultimately, we utilized a dataset comprising 993 benign applications and 991 malicious applications.

\subsection{ARA Technology Detection}
\label{sec:ARA Technology Detection}

\vspace{0.5em}

Detecting the use of ARA technology in such a large-scale dataset swiftly and accurately poses a significant challenge in this study. ARA technology encompasses a wide range of types and complex implementations, making it a non-trivial task to conduct rapid and precise assessments.

In this study, we adopt a systematic and representative categorization of ARA techniques guided by the Mobile Application Security Verification Standard (MASVS), whose resiliency requirements cover common runtime tampering and anti-analysis mechanisms. Following these requirements—and to make the taxonomy practical for large-scale empirical evaluation—we consolidate the landscape into five representative categories (AD, AT, AH, RD, VED) that are widely deployed in real-world apps and well supported by our detector. These five categories serve as the basis for our measurements and analysis across the dataset.

To operationalize this classification, we employed ARAP~\citep{Suo2024ARAPDA} for ARA detection. ARAP combines static and dynamic analysis to identify concrete ARA techniques via feature matching. It maintains a list of 1{,}515 distinct features and supports detection across the above five categories and 32 finer-grained subcategories. This enables us to consistently map observed defenses in APKs to the five-category taxonomy used throughout the paper.

Due to space limitations, the detailed implementation and workflow of ARAP are provided in our prior work, where extensive experiments demonstrate high detection accuracy. We also conducted a comparative analysis with ATADetector~\citep{Berlato2020ALS}—the only other available tool in this area—and found that ARAP significantly outperforms ATADetector in overall detection performance.

The output of ARAP is a JSON report that records the various ARA techniques identified. This report allows for easy identification of the number of different ARA techniques used in a specific APK. Based on the number of ARA techniques detected, we classified the APKs into four categories: EASY (0 ARA techniques), NORMAL (1–5 ARA techniques), HARD (6–10 ARA techniques), and CHALLENGING (more than 10 ARA techniques).

\vspace{-1em}
\subsection{Code Coverage Measurement}
\label{sec:Code_Coverage_Measurement}

Another challenge faced in our research is achieving reliable fine-grained code coverage measurement for closed-source APKs when access to the source code is unavailable.  Achieving reliable code coverage in a black-box setting is a non-trivial task.

In our study, we utilized ACVTool~\citep{pilgun2020acvtool} as the code coverage detection tool. ACVTool measures fine-grained coverage for third-party Android applications by inserting probes into the smali representation of Dalvik bytecode, supporting class-, method-, and instruction-level granularity. Following the tool authors’ recommendation, we adopt the default \emph{instruction-level} coverage and report instruction coverage in all our experiments. ACVTool can be integrated with any testing or dynamic analysis framework, which fits our experimental pipeline. According to the original authors, ACVTool introduces negligible instrumentation overhead and exhibits very high reliability.

Prior to conducting the experiments, we used ACVTool to instrument all APKs to ensure that ACVTool could work properly on these APKs. The final instrumentation results, as shown in Table \ref{tab:acvtool Instrumentation Results}, indicated that 81.4\% of them were successfully instrumented by ACVTool. We conducted subsequent experiments using these 1,615 instrumented APK files.

\begin{table}[h]
\centering
\caption{ACVTool Instrumentation Results}
\renewcommand{\arraystretch}{1.25}
\label{tab:acvtool Instrumentation Results}
\begin{tabular}{ccc}
\toprule
\textbf{Category} & \textbf{Success Rate (\%)} & \textbf{Success Count} \\
\midrule
Benign & 86.3 & 857 \\
Malicious & 76.5 & 758 \\
Overall & 81.4 & 1,615\\
\bottomrule
\end{tabular}
\label{tab:installation_stats}
\vspace{-1em}
\end{table}

\vspace{-1em}
\subsection{Dynamic Analysis Tools}
 To conduct a comprehensive evaluation of dynamic analysis tools and ensure that we covered as many relevant tools as possible, we performed searches using keywords such as "Android," "Dynamic," "Analysis," and "Tool" in well-known literature databases, including IEEE Xplore, ACM Digital Library, Springer Link, and ScienceDirect. Furthermore, we manually reviewed the papers to identify those that included active tools, with particular attention to the "Background" and "Related Work" sections to ensure no relevant tools were overlooked.

 As a result, we reviewed 58 articles published from 2012 to the present that discussed these tools, identifying 65 distinct tools. After this review, we attempted to replicate the identified tools and excluded those that could not be built due to issues such as lack of support, incomplete build information, or incompatibility with our operational analysis environment (Section \ref{sec:Evaluation_Framework}). This process led to the confirmation of 6 usable dynamic analysis tools, as shown in Table \ref{tab:Dynamic analysis tools}. For transparency and reproducibility, we provide a comprehensive list of all reviewed tools, including excluded ones and reasons for exclusion, in our open-source repository, along with detailed reasons for their exclusion.


\begin{table}[]
\centering
\caption{Dynamic Analysis Tool Are Involved in This Study}
\renewcommand{\arraystretch}{1.25}
  \label{tab:Dynamic analysis tools}
  \renewcommand{\arraystretch}{1.2} %
\begin{tabular}{ccc}
\toprule
\textbf{Tool}  & \textbf{Year} & \textbf{Function}        \\ \toprule
APIMonitor     & 2013          & Dynamic Behavior Capture \\ 
AndroidSlicer  & 2019          & Dynamic Slicing          \\ 
DroidCat       & 2019          & Dynamic Behavior Capture \\ 
T-Recs         & 2022          & Dynamic Taint Analysis   \\ 
DroidDissector & 2023          & Dynamic Behavior Capture \\ 
ESdroid        & 2023          & Dynamic Slicing          \\ \bottomrule
\end{tabular}
\vspace{-1.5em}
\end{table}

$\bullet$~\textbf{APIMonitor}~\citep{APIMonitor} inserts monitoring code into the APK file.  By running the repackaged APK, APIMonitor can capture API call logs to understand the behavior of the application.  The analysis includes information on inbound/outbound network data, file read/write operations, and activities such as sending messages and making calls.

Although APIMonitor was released in 2013, it remains relevant to our study for several reasons.  Firstly, we aimed for a comprehensive review and analysis of dynamic analysis tools, ensuring we did not overlook any potential tools that might contribute to the current landscape.  Secondly, APIMonitor claims to fully consider backward compatibility, making it suitable for analyzing modern applications despite its age.  Thus, its inclusion in our study is justified, as it provides valuable insights into dynamic analysis, particularly in terms of API-level monitoring.

$\bullet$~\textbf{AndroidSlicer}~\citep{Azim2019DynamicSF} combines a novel asynchronous slicing approach for modeling data and control dependencies in the presence of callbacks with lightweight and precise instrumentation; this allows slicing for apps running on actual phones, and without requiring the app's source code.

$\bullet$~\textbf{DroidCat}~\citep{Cai2019DroidCatEA} is a dynamic app classification tool that uses a diverse set of dynamic features based on method calls and inter-component communication (ICC) Intents. It does not rely on permissions, app resources, or system calls and can fully handle reflection. This approach achieves superior robustness compared to static methods and dynamic approaches that rely on system calls.

$\bullet$~\textbf{T-Recs}~\citep{Inayoshi2022PlugAA} is a novel dynamic taint analyzer.  It tracks information flow by recording application bytecode-level executions on Android devices and reconstructing the executions on a server independent of specific Android versions and devices.  

$\bullet$~\textbf{DroidDissector}~\citep{Muzaffar2023DroidDissectorAS} is a tool for extracting static and dynamic features.  It aims to provide malware researchers and analysts in the Android domain with an integrated tool capable of extracting all commonly used features in Android malware detection.  Its dynamic analysis module analyzes the complete behavior of applications by tracking system calls, generated network traffic, API calls, and generated log files used by the application.

$\bullet$~\textbf{ESdroid}~\citep{Win2023EventawarePD} is a dynamic slicing technique that is event-aware and suitable for Android applications.  The novelty of ESdroid lies in its combination of segment-based incremental debugging and dynamic backward slicing to narrow down the search space, thereby generating precise slices for Android.

\subsection{Evaluation Framework}
\label{sec:Evaluation_Framework}
\begin{figure}[t]
\centerline{\includegraphics[width=1.0\linewidth]{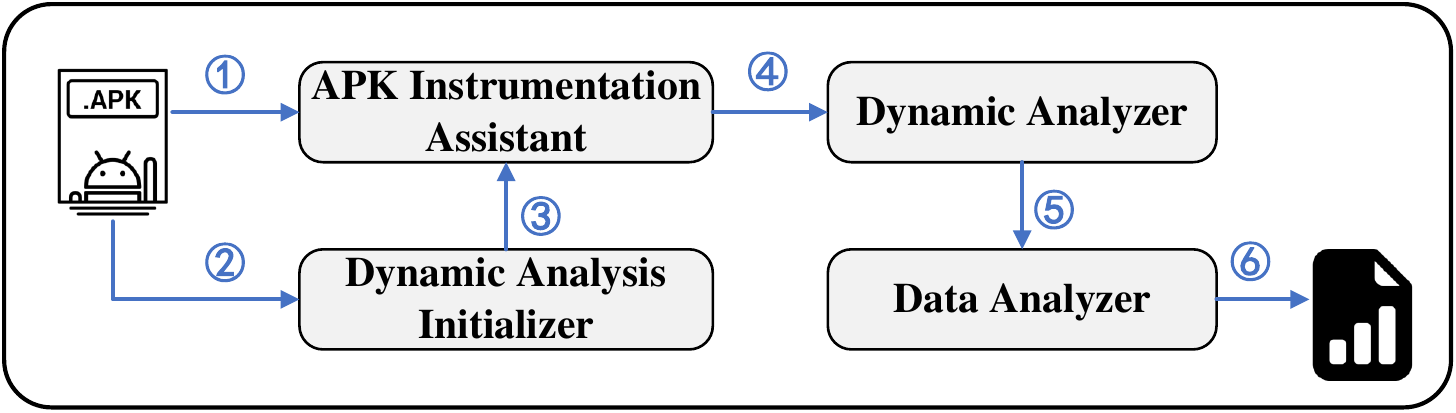}}
\caption{Study Methodology}
\label{fig:Framework}
\vspace{-1.5em}
\end{figure}

To conduct our study, we developed a modular evaluation framework (Figure \ref{fig:Framework}), consisting of four key components: APK Instrumentation Assistant, Dynamic Analysis Initializer, Dynamic Analyzer, and Data Analyzer. Our framework is designed to be reusable and easily extendable, allowing users to integrate new dynamic analysis tools for performance assessment.  In the spirit of transparency and collaboration, we have decided to open-source this tool~\citep{Anti-ARA}. The framework is written in Python and consists of over 3,300 lines of code.

In our framework, there are two distinct paths, referred to as Route 1 (2, 3, 4, 5, 6) and Route 2 (1, 4, 5, 6).     This is because we want to demonstrate the capabilities of dynamic analysis tools by comparing the changes in code coverage after processing the APK code through these tools.     In Route 2, the APK is only processed by ACVTool, which serves as a baseline, reflecting code coverage achievable through dynamic analysis without modifications to ARA techniques.  In Route 1, the APK is first preprocessed by dynamic analysis tools, which may bypass certain ARA techniques, thus potentially resulting in higher code coverage.     By comparing the difference in code coverage between Route 1 and Route 2, we can evaluate the capabilities of dynamic analysis tools.

\textbf{APK Instrumentation Assistant.} This module takes an Android APK file as input, utilizing ACVTool to statically instrument the APK for the purpose of obtaining code coverage during dynamic analysis.  Additionally, in order to better trigger ARA technology, the APK Instrumentation Assistant will add "android:debuggable="true"" to the APK's manifest file, setting it as debuggable to trigger AD technology. At the same time, the repackaged APK will help trigger the AT technology.

\textbf{Dynamic Analysis Initializer.} In Route 2, the APK will first be processed by this module. This module invokes various dynamic analysis tools to preprocess the input APK file and generate the processed APK file. We expect that the APK, after being preprocessed by dynamic analysis tools, will have enhanced analyzability, thereby achieving higher code coverage.

To further clarify this logic, when an APK is analyzed dynamically, the application may attempt to detect certain runtime anomalies, such as the presence of debugging tools. In its original state, upon detecting such anomalies, the application might take actions to prevent further analysis, such as displaying a message to the user and terminating the program, or restricting certain features to avoid detection. These self-defense mechanisms limit the analysis process and result in lower code coverage.

However, when dynamic analysis tools preprocess the APK by inserting monitoring and logging code, they modify the application's original behavior. This alteration reduces the application's capacity to detect debugging environments or other runtime anomalies. Consequently, even if the application recognizes an abnormal runtime environment, it can no longer take defensive actions like terminating itself or blocking certain functionalities. As a result, the instrumented version of the APK becomes more analyzable, and a larger portion of the application’s functionality is explored during testing, leading to higher code coverage compared to the original APK.

This enhanced analyzability and increased code coverage in the instrumented APK are direct results of the dynamic analysis tools modifying the application’s logic, which prevents the application from evading analysis through runtime detection mechanisms.

\textbf{Dynamic Analyzer.} This module loads the input APK file into an emulator for dynamic analysis and utilizes ACVTool to capture the corresponding code coverage.  In this study, we choose the Android Virtual Device as the emulator environment.  To achieve automated application exploration, we use the well-known automation testing tool Monkey~\citep{Monkey}. For each dynamic analysis, we use Monkey to randomly generate 50,000 events as input.  For each APK file, we conduct 3 analyses and take the average of the 3 runs as the final code coverage.  After each dynamic analysis, we use snapshots to restore the emulator's state to ensure the independence of each analysis. For each set of controlled experiments (Route 1 and Route 2), we used the same -s parameter to generate identical event sequences, thereby minimizing any bias introduced by the Monkey tool.

Furthermore, in this study, effective triggering of ARA techniques embedded in the APK is essential. Assuming that the ARA technology used in the APK is not triggered, the research process becomes meaningless because it is equivalent to the APK not utilizing any ARA technology. To address this issue, we meticulously designed the dynamic analysis environment in the Dynamic Analyzer to maximize the triggering of the ARA technology used in the application. Specifically, in addition to using the emulator as the runtime environment to trigger VED technology, we also 1) set the emulator state to root to trigger RD technology; 2) configure 
Xposed and Frida in the emulator to trigger AH technology.

\textbf{Data Analyzer.} This module analyzes and aggregates the results from both evaluation routes, storing them in a CSV file and generating statistical reports. These reports enable users to easily assess and compare the performance of the dynamic analysis tools under evaluation.

Based on the methods described above, we conducted controlled experiments to assess the performance of various dynamic analysis tools in handling ARA techniques. In the next section, we present and analyze the experimental results, focusing on key metrics such as code coverage and efficiency, and examining how different ARA techniques impact dynamic analysis.

\vspace{-1em}
\section{DATA ANALYSIS AND RESULTS}
\label{sec:DATA_ANALYSIS_AND_RESULTS}

In the previous section, we outlined the methodology used to evaluate the effectiveness of dynamic analysis tools in addressing Anti-Runtime Analysis (ARA) technologies.   In this section, we present the results of the experiments based on the outlined methodology.   Our focus is on key metrics such as code coverage and the efficiency of dynamic analysis tools.

To carry out the experiments, we utilized four desktop computers for parallel analysis.   Each system was equipped with an Intel Xeon E-2224G processor and 16GB of memory, running Android version 12 on the Android Virtual Device.   The entire experiment was conducted over a span of 3 months.   The following subsections will present and analyze the results obtained from these experiments.

\vspace{-0.3em}
\newtcolorbox[auto counter, number within=section, list type=subsubsection, list inside=toc]{sectionbox}[2]{
    colback=white, colframe=black,
    colbacktitle=white!80!gray, coltitle=black,
    fonttitle=\bfseries, title={#1}, list entry={#1 \thetcbcounter: #2\quad},
    before upper={\parindent10pt\noindent},
    left = 1mm,
    right = 1mm,
    top = 1mm,
    bottom = 1mm,
}
\vspace{-0.3em}

\newtcolorbox[auto counter, number within=section, list type=subsubsection, list inside=toc]{sectionboxN}[2]{
    colback=white, colframe=black,
    colbacktitle=white!80!gray, coltitle=black,
    fonttitle=\bfseries, title={#1}, list entry={#1 \thetcbcounter: #2\quad},
    before upper={\parindent10pt\noindent},
    coltext=blue, 
    left = 1mm,
    right = 1mm,
    top = 1mm,
    bottom = 1mm,
}

\newcommand{\rcomment}[2]{
    \begin{sectionbox}{#1}
        #2
    \end{sectionbox}
}

\newcommand{\rcommentN}[2]{%
    \begin{sectionboxN}{#1}%
        #2
    \end{sectionboxN}%
}

\begin{tcolorbox}[left=2mm, right=2mm, top=1mm, bottom=1mm]
\faLightbulbO \ \  {\textbf{Q1: Can dynamic analysis tools effectively handle ARA technology?}}
\end{tcolorbox}

In the first research question, we aim to evaluate the ability of various dynamic analysis tools to handle ARA techniques from a holistic perspective. Ideally, such tools should possess certain capabilities to mitigate the impact of ARA techniques. Under Route 2—where no preprocessing or instrumentation is applied—the APK under analysis retains all its ARA mechanisms, which can be triggered by our specially crafted runtime environment. As a result, dynamic analysis under Route 2 generally yields lower code coverage due to the active ARA techniques. In contrast, Route 1 introduces preprocessing steps by the dynamic analysis tools prior to execution. This may neutralize or bypass some ARA techniques, potentially resulting in higher code coverage under identical input conditions.

\begin{figure*}[t]
\centerline{\includegraphics[width=1.0\linewidth]{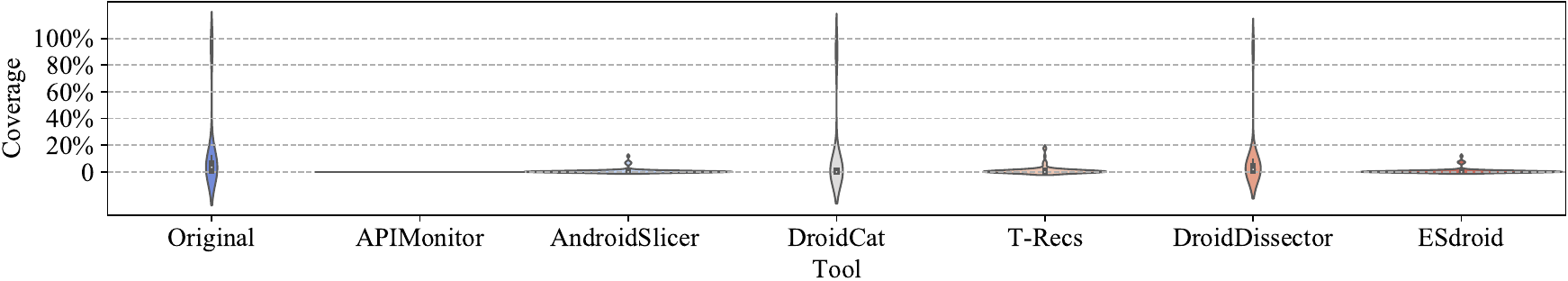}}
\vspace{-1em}
\caption{Code Coverage of the Malicious APKs}
\label{fig:Code coverage of the malicious apk}
\vspace{-1em}
\end{figure*}

\begin{figure*}[t]
\centerline{\includegraphics[width=1.0\linewidth]{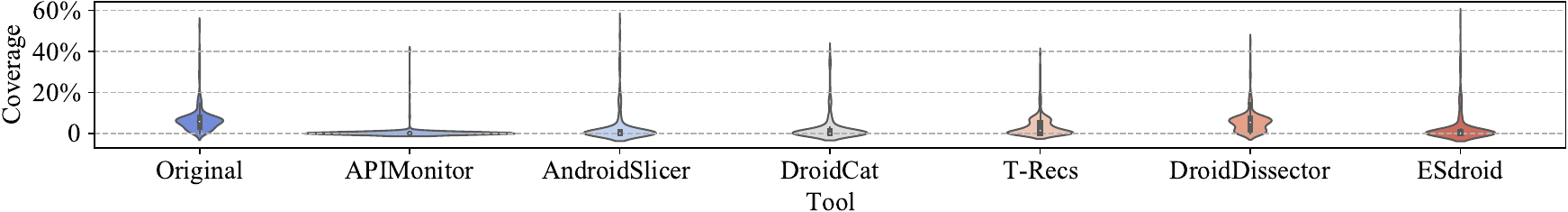}}
\vspace{-1em}
\caption{Code Coverage of the Benign APKs} 
\label{fig:Code coverage for benign apk}
\vspace{-1.5em}
\end{figure*}

Figures \ref{fig:Code coverage of the malicious apk} and \ref{fig:Code coverage for benign apk} respectively illustrate the code coverage during dynamic analysis of malicious and benign APKs. The results of six different dynamic analysis tools are compared with the analysis results of the original, untreated APKs on the same coordinate system, facilitating a clearer comparison of the performance of each tool. The "Original" results in Figures \ref{fig:Code coverage of the malicious apk} and \ref{fig:Code coverage for benign apk} correspond to the experimental configuration described as Route 2 in Figure \ref{fig:Framework}, where no preprocessing is performed and ARA techniques remain active during execution. In the violin plots within Figures \ref{fig:Code coverage of the malicious apk} and \ref{fig:Code coverage for benign apk}, the y-axis represents the code coverage achieved during dynamic analysis, while the width of the violin plot indicates the density of samples at a given coverage level. Note that the filled area appearing below the axis in the violins does not indicate negative coverage values; it is a rendering artifact caused by the presence of zero-coverage samples and the polygon closure used by the plotting library. In all analyses, an app–tool configuration that fails to produce a valid run (instrumentation failure, install/startup crash, emulator deadlock/timeout) is assigned 0\% coverage, reflecting the tool’s effective runtime visibility for that app.

By comparing the results for benign and malicious APKs, we observe that malicious APKs achieve higher code coverage than benign APKs when the same number of inputs (50,000) are provided. This outcome is expected, given that benign APKs typically possess broader functionality and larger codebases, which result in lower code coverage during dynamic analysis. Specifically, in our dataset, malicious APKs have an average code size of 6.09 MB, whereas benign APKs average 10.98 MB. To eliminate the impact of resource files on size measurements, we calculated only the size of the application code.

More critically, when comparing the "Original" baseline to the results obtained after applying each dynamic analysis tool (i.e., Route 2 vs. Route 1), we observe no significant improvement in code coverage—regardless of whether the APK is benign or malicious. This indicates that none of the tested dynamic analysis tools effectively handle the ARA techniques deployed in these applications. It is this comparison—between the tools-processed results and the baseline scenario—that leads us to conclude the current ineffectiveness of dynamic analysis tools in overcoming runtime ARA defenses.

For malicious APKs, DroidCat and DroidDissector are the two best-performing tools. However, "best-performing" only means that their analysis results are closest to the original results. The code coverage during dynamic analysis of the APKs slightly decreased after being processed by these tools. For the other four tools—APIMonitor, AndroidSlicer, T-Recs, and ESdroid—the code coverage during dynamic analysis is significantly lower than that of the unprocessed APKs. In particular, APIMonitor renders the processed APK files nearly unanalyzable.

The performance of the dynamic analysis tools on benign APKs was similarly disappointing, with no tool effectively handling ARA techniques to improve code coverage during dynamic analysis. However, unlike the results for malicious APKs, all the tools performed somewhat better when analyzing benign APKs. Among them, DroidDissector performed the best, with its code coverage nearly identical to the original results. This was followed by T-Recs, which slightly reduced the APK's code coverage. AndroidSlicer, DroidCat, and ESdroid displayed similar performance, all significantly reducing the APK's code coverage. Although APIMonitor performed better on benign APKs than on malicious ones, it still rendered most APKs unanalyzable, reducing the code coverage to zero.

The experimental results for Q1 were surprising.   Not only did the code coverage during dynamic analysis of APKs processed by dynamic analysis tools fail to improve, but only a few tools could ensure that code coverage during dynamic analysis did not significantly decline.

Among all tools, APIMonitor demonstrated the poorest performance. APKs processed by APIMonitor became nearly unanalyzable.   This poor performance is partly attributable to the age of APIMonitor;   it cannot effectively handle newer APK files.   Although APIMonitor's authors claim that it has some backward compatibility and can work on newer versions of Android, our experimental results indicate that it struggles to function effectively on more recent Android versions.

DroidCat and DroidDissector, like APIMonitor, are Dynamic Behavior Capture tools and were the best performers on the malicious APK dataset. This suggests that, compared to other types of tools, these two were more designed with the analysis of malicious APKs in mind. DroidDissector also performed well on the benign dataset, whereas DroidCat's performance declined on it. T-Recs, which is used for taint analysis, did not perform well on the malicious APK dataset, indicating it is more suitable for analyzing benign APKs. The two Dynamic Slicing tools, AndroidSlicer and ESdroid, performed poorly on the malicious APK dataset but showed better results on the benign dataset, indicating they are more appropriate for analyzing benign APKs.

\begin{table*}[t]
\centering
\caption{Paired Wilcoxon tests (tool vs.\ \textit{Original}) with Holm correction on benign and malicious datasets. 
The rank-biserial correlation ($r$) indicates effect size (negative = lower coverage than baseline). 
All results are significant at $p_{\text{holm}} < 0.05$.}
\label{tab:wilcoxon_combined}
\begin{tabular}{lcccccc}
\toprule
\multirow{2}{*}{\textbf{Tool}} 
& \multicolumn{3}{c}{\textbf{Benign Dataset}} 
& \multicolumn{3}{c}{\textbf{Malicious Dataset}} \\
\cmidrule(lr){2-4} \cmidrule(lr){5-7}
 & \textbf{$r$} & \textbf{$p_{\text{holm}}$} & \textbf{Sig.} 
 & \textbf{$r$} & \textbf{$p_{\text{holm}}$} & \textbf{Sig.} \\
\midrule
APIMonitor     & $-0.972$ & $2.18\times10^{-15}$ & \textbf{TRUE} & $-0.991$ & $1.30\times10^{-16}$ & \textbf{TRUE} \\
AndroidSlicer  & $-0.778$ & $4.25\times10^{-12}$ & \textbf{TRUE} & $-0.836$ & $2.52\times10^{-12}$ & \textbf{TRUE} \\
ESdroid        & $-0.758$ & $7.10\times10^{-12}$ & \textbf{TRUE} & $-0.813$ & $5.09\times10^{-12}$ & \textbf{TRUE} \\
T-Recs         & $-0.721$ & $3.85\times10^{-10}$ & \textbf{TRUE} & $-0.781$ & $1.46\times10^{-10}$ & \textbf{TRUE} \\
DroidCat       & $-0.503$ & $1.09\times10^{-6}$  & \textbf{TRUE} & $-0.579$ & $8.47\times10^{-7}$  & \textbf{TRUE} \\
DroidDissecto  & $-0.235$ & $3.55\times10^{-2}$  & \textbf{TRUE} & $-0.271$ & $4.00\times10^{-2}$  & \textbf{TRUE} \\
\bottomrule
\end{tabular}
\vspace{-1.5em}
\end{table*}

To further validate the robustness of our findings, we conducted paired Wilcoxon signed-rank tests to compare each dynamic analysis tool with the baseline (\textit{Original}). 
The Wilcoxon test was chosen over parametric alternatives because code-coverage data are non-normally distributed, bounded within [0,1], and contain many zero values, 
making non-parametric methods more appropriate and robust. 
Each test was performed on paired coverage values from the same APK before and after instrumentation, ensuring direct comparability. 
To account for multiple comparisons across tools, we applied the Holm correction to control the family-wise error rate while maintaining higher statistical power than the Bonferroni adjustment. 
Additionally, we computed the rank-biserial correlation ($r$) as an effect-size measure, which complements $p$-values by indicating both the direction and magnitude of change.

We performed this statistical analysis separately on the benign and malicious datasets to account for potential behavioral differences between the two categories. 
Across both datasets, all six tools exhibited statistically significant differences from the baseline after Holm correction ($p_{\text{holm}} < 0.05$), with negative $r$ values for all tools, as summarized in Table~\ref{tab:wilcoxon_combined}. 
This pattern indicates that, in general, dynamic analysis tools achieved lower code coverage than the baseline configuration, suggesting that ARA defenses remain effective in impeding runtime exploration. 
The effect sizes were slightly stronger in the malicious dataset (e.g., \textit{APIMonitor}: $r \approx -0.99$; \textit{AndroidSlicer}: $r \approx -0.84$) than in the benign dataset (e.g., \textit{APIMonitor}: $r \approx -0.97$; \textit{AndroidSlicer}: $r \approx -0.78$), 
which aligns with the expectation that malicious applications tend to implement more complex and aggressive ARA protections. 
These results quantitatively confirm the conclusions drawn from our coverage-based evaluation: existing dynamic analysis tools remain substantially limited in their ability to counter ARA mechanisms, particularly when analyzing malware.

\vspace{-0.3em}
\rcomment{Q1 Finding}{ -We discovered that the existing dynamic analysis tools are generally ineffective at handling the ARA techniques used in APKs.  Among them, DroidDissector performed the best, with its results on both the benign and malicious datasets closely matching the original results. }

\vspace{-0.3em}

\begin{tcolorbox}[left=2mm, right=2mm, top=1mm, bottom=1mm]
\faLightbulbO \ \  {\textbf{Q2: What are the differences in the impact of various categories and quantities of ARA technology on dynamic analysis tools?}}
\end{tcolorbox}

In Question 2, we aim to explore how different types and quantities of ARA techniques impact the performance of dynamic analysis tools, specifically their effectiveness in achieving code coverage. We begin by analyzing the impact of ARA technique categories on dynamic analysis outcomes.

We combined the benign and malicious APK datasets from Question 1 and categorized them based on whether they employed a specific category of ARA technique. In other words, an APK that deploys multiple different categories of ARA techniques would be placed into multiple categories. To assess the impact of ARA technology categories on code coverage during dynamic analysis, we focused on the five major categories into which ARAP divides ARA techniques.

\begin{figure*}[h]

\centerline{\includegraphics[width=1.0\linewidth]{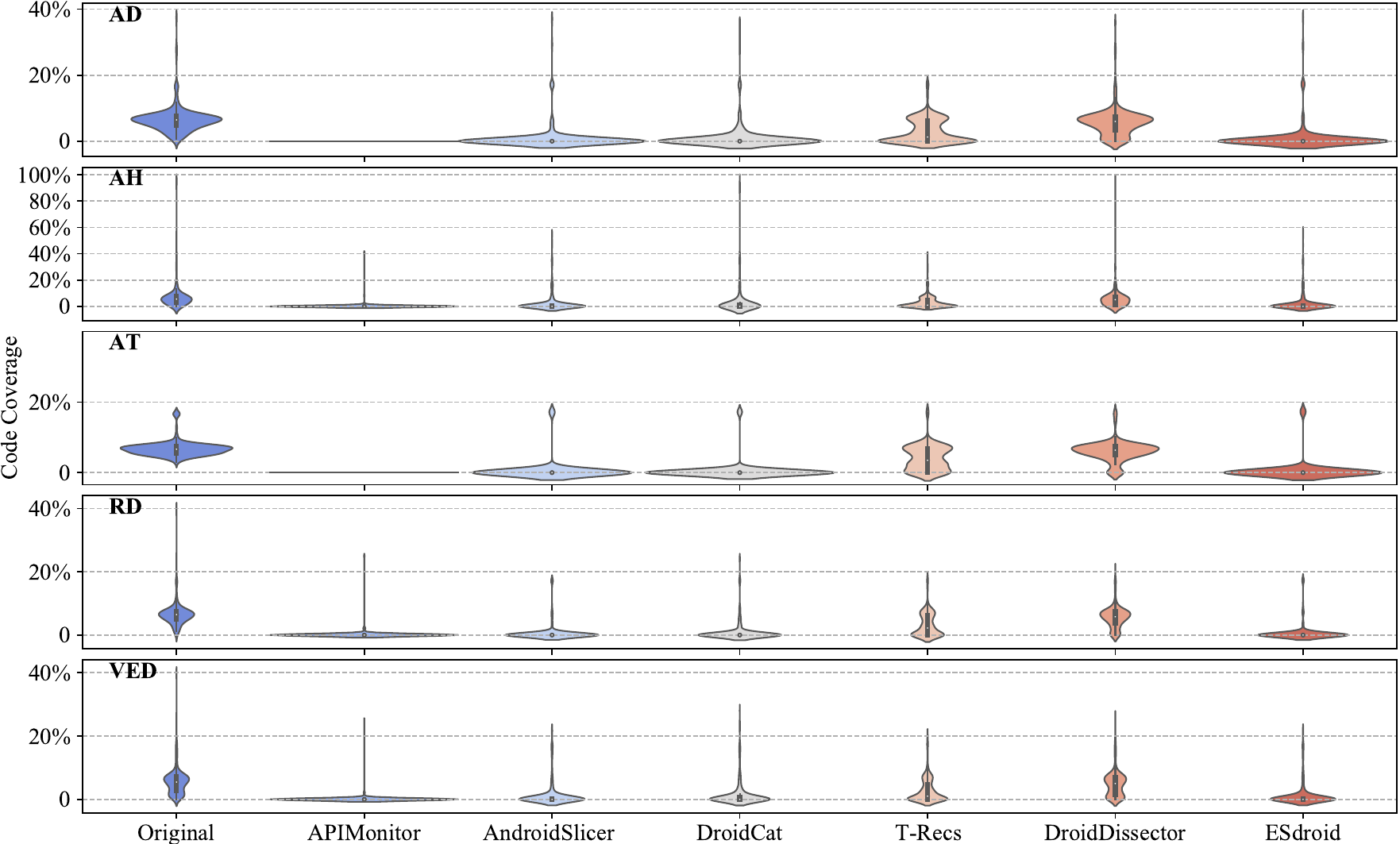}}
\vspace{-1em}
\caption{Code Coverage Rates for Different Categories of ARA techniques}
\label{fig:Code Coverage by Different ARA Categories}
\vspace{-1.5em}
\end{figure*}

Figure \ref{fig:Code Coverage by Different ARA Categories} presents our experimental results, with five subfigures displayed top to bottom, each showcasing the outcomes of APK datasets incorporating the AD, AH, AT, RD, and VED techniques. By observing Figure \ref{fig:Code Coverage by Different ARA Categories} , we have drawn some interesting conclusions. 

\begin{table}[]
\caption{Average Mean and Average Median of Code Coverage Rates under Different Technologies}
    \centering
    \renewcommand{\arraystretch}{1.25}  
    \begin{tabular}{ccc}
        \toprule
         \textbf{Technology} & \textbf{Average Mean (\%)} & \textbf{Average Median (\%)}  \\
        
        \midrule
        AD                  & 8.61                                 & 8.66                                    \\
        AH                  & \textcolor{red}{9.21}                                  & \textcolor{red}{9.19}                                   \\
        AT                  & 8.73                                  & 8.65                                    \\
        RD                  & 8.43                                  & 8.45                                   \\
        VED                 & \textcolor{orange}{7.49}                                  & \textcolor{orange}{7.47}                                    \\
        \bottomrule
    \end{tabular}
    \label{tab:Average Mean and Average Median of Code Coverage Rates under Different Technologies}
    \vspace{-1.5em}
\end{table}

To more intuitively demonstrate the impact of different categories of ARA techniques on code coverage, we calculated the average mean and average median values of the six tools under the influence of various ARA techniques (Table \ref{tab:Average Mean and Average Median of Code Coverage Rates under Different Technologies}). In the table, the highest values for each metric are highlighted in red, and the lowest values are highlighted in orange. In this document, subsequent tables also adhere to this convention.  

By comparing the experimental results of the original APK set, we observed that dynamic analysis tools achieved the lowest code coverage on APKs deploying VED techniques among the five categories. This observation suggests a potential association between VED and reduced code exposure, possibly due to its ability to detect virtual execution environments and alter application behavior accordingly. Conversely, tools achieved relatively higher code coverage on APKs deploying AH, with some samples even exceeding 90\%. This may imply that AH, as implemented in these samples, is comparatively less effective in hindering dynamic analysis.

It is important to note that dynamic analysis outcomes can be influenced by multiple factors, including the randomness of input event generation and the intrinsic variability across application sets associated with each ARA technique. As previously described in Section~\ref{sec:Research_Methodology}, the experimental design leverages a large-scale dataset of 1,615 applications, with each APK undergoing three independent runs of 50,000 Monkey-generated events. This setup helps reduce the impact of nondeterministic UI behavior and supports more statistically robust comparisons across different ARA technique categories.

The relatively high coverage in AH-deploying apps may also reflect characteristics of real-world deployment: unlike other techniques, AH lacks formal mitigation support in Android and is known for its implementation complexity, which may limit its adoption or lead to inconsistent enforcement. These factors collectively suggest that AH may be less reliably implemented in the wild, contributing to its limited efficacy in practice.
\vspace{0.2em}
In contrast, VED has been shown in prior research (e.g., ARAP) to be widely adopted—appearing in 92.6\% of benign apps and 33.0\% of malicious apps—likely due to its native support in the Android ecosystem. VED is also known to be adopted across a broad spectrum of application domains, suggesting that the observed reduction in code coverage is primarily attributable to the mechanism of the VED technique itself rather than category-specific application behaviors.

Moreover, it was observed that none of the samples in the AT-deploying group exceeded 20\% code coverage during dynamic analysis. While multiple factors may contribute to this result, one plausible explanation relates to the nature of AT implementations. According to findings from ARAP, AT techniques are often deployed using Android-supported official mechanisms such as signature verification and installer checks, which are typically executed at early stages of the application lifecycle. This design increases the likelihood of early application termination upon detection of tampering, potentially limiting dynamic execution paths. Furthermore, the widespread use of these standardized mechanisms may discourage the adoption of more complex or customized AT logic, reducing the chance that low code coverage results from application complexity alone.

The AD and RD techniques demonstrated similar behavior, with most samples exhibiting coverage levels below 40\%, suggesting that these mechanisms can also effectively constrain dynamic code exploration, albeit to a lesser degree.

For various dynamic analysis tools, APIMonitor, AndroidSlicer, DroidCat, and ESdroid show minimal differences in their performance across the five categories of ARA techniques, without any of them exhibiting a clear inclination towards handling a particular category of technique.  In contrast, T-Recs and DroidDissector outperform the other four tools but do not demonstrate a marked preference for handling any specific category technique.

\begin{table}[t]
\centering
\caption{Pairwise differences in median coverage between ARA categories using app-level cluster bootstrap (B=10{,}000). Stars reflect Holm-adjusted significance: *** $p<0.001$, ** $p<0.01$, * $p<0.05$, (n.s.) not significant.}
\label{tab:ara_category_pairwise_bootstrap}
\scriptsize
\setlength{\tabcolsep}{4pt}
\begin{tabular}{l l @{\hspace{1.2em}} l l}
\toprule
\multicolumn{2}{c}{\textbf{Non-VED contrasts}} & \multicolumn{2}{c}{\textbf{VED contrasts}} \\
\cmidrule(lr){1-2}\cmidrule(lr){3-4}
\textbf{Pair} & \textbf{$\Delta$ median [pp] (95\% CI)} & \textbf{Pair} & \textbf{$\Delta$ median [pp] (95\% CI)} \\
\midrule
AD--AH  & 0.87 \,[0.62, 1.05]$^{***}$  & AD--VED & 0.98 \,[0.75, 1.18]$^{***}$ \\
AD--AT  & $-0.18$ \,[$-0.43$, 0.05] (n.s.) & AT--VED & 1.17 \,[0.89, 1.36]$^{***}$ \\
AD--RD  & 0.02 \,[$-0.13$, 0.17] (n.s.) & AH--VED & 0.11 \,[0.03, 0.20]$^{**}$ \\
AT--AH  & 1.06 \,[0.75, 1.29]$^{***}$  & RD--VED & 0.94 \,[0.71, 1.13]$^{***}$ \\
AT--RD  & 0.22 \,[0.05, 0.42] (n.s.)    &         & \\
AH--RD  & $-0.83$ \,[$-1.05$, $-0.65$]$^{***}$ &         & \\
\bottomrule
\end{tabular}
\vspace{-1.5em}
\end{table}

To formally support the observations in Table~\ref{tab:Average Mean and Average Median of Code Coverage Rates under Different Technologies}, we performed an app-level \emph{cluster bootstrap} (B=10{,}000): apps were resampled with replacement, median coverage was recomputed per category for each resample, and all pairwise contrasts were formed; percentile 95\% CIs and two-sided $p$-values were obtained from the empirical distributions with Holm correction across the ten contrasts. As summarized in Table~\ref{tab:ara_category_pairwise_bootstrap}, the contrasts that involve VED are consistently significant and indicate lower median coverage for VED than for the other categories, quantitatively reinforcing that VED imposes the strongest resistance to dynamic analysis. Among the remaining categories (AD/AT/AH/RD), pairwise differences are comparatively small; where significance occurs, the effect sizes are modest and do not alter the qualitative ordering shown in Table~\ref{tab:Average Mean and Average Median of Code Coverage Rates under Different Technologies}.

\begin{figure*}[h]
\centerline{\includegraphics[width=1.0\linewidth]{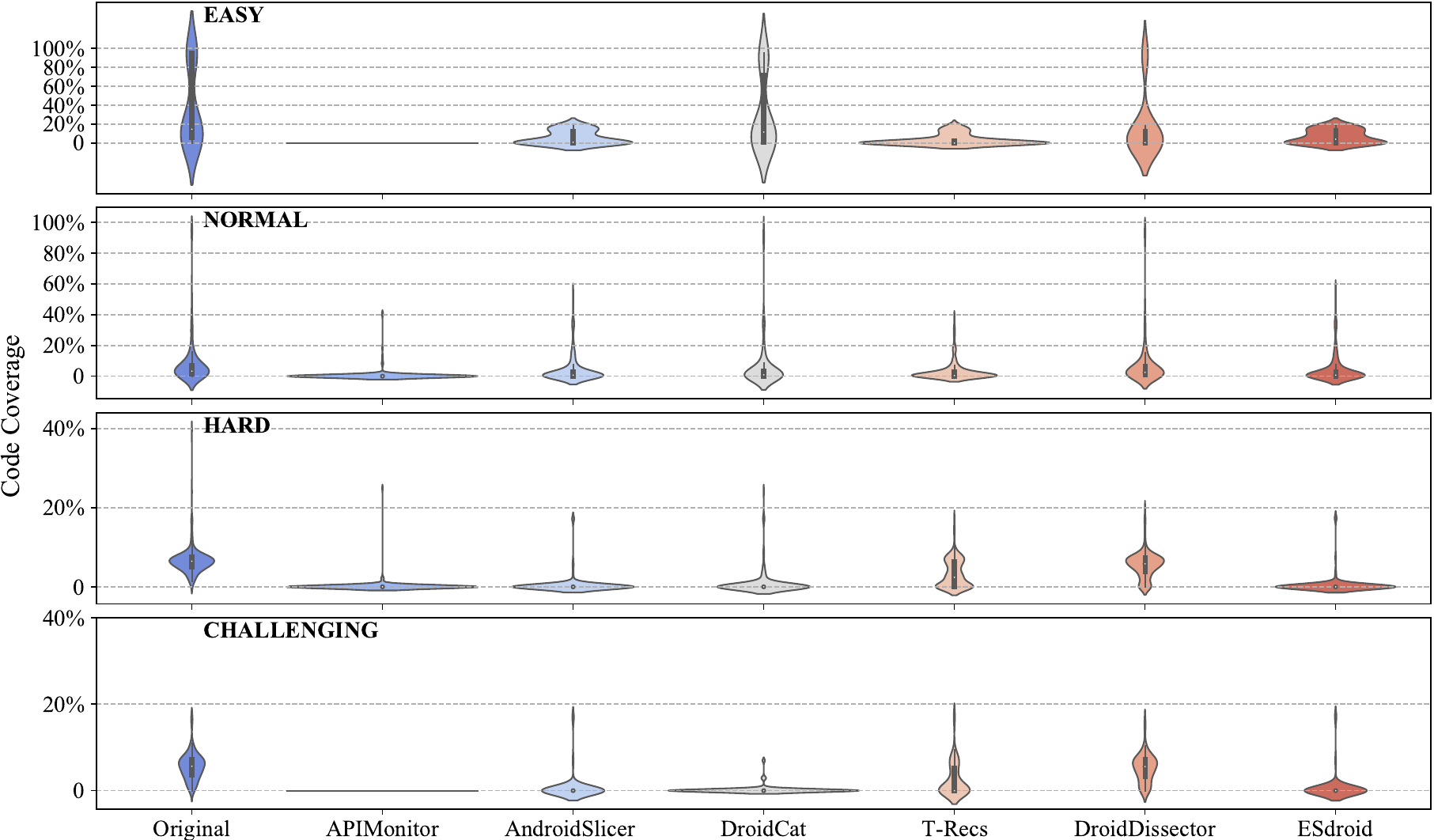}}
\caption{Code Coverage Rates for Different Quantity of ARA techniques} 
\vspace{-1em}
\label{fig:Code Coverage by Different ARA Quantity}
\vspace{-1.5em}
\end{figure*}

In Question 2, the second aspect we need to discuss is the impact of the number of ARA techniques deployed on the code coverage achievable during dynamic analysis. Figure \ref{fig:Code Coverage by Different ARA Quantity} illustrates our experimental results. As described in Section \ref{sec:ARA Technology Detection}, we categorized the APKs into four datasets—EASY, NORMAL, HARD, and CHALLENGING—based on the number of ARA techniques employed, with an increasing number of techniques used for each subsequent category. When discussing the impact of the number of ARA techniques on the code coverage achievable during dynamic analysis, we referred to the subcategory information from the outputs of the ARAP tool, dividing ARA techniques into 32 subcategories. The distribution of samples across these four subsets is as follows: EASY (22), NORMAL (730), HARD (687), and CHALLENGING (176). The relatively small number of EASY and CHALLENGING samples reflects an earlier observation: few applications completely avoid ARA techniques, particularly those released after 2020, while excessive use of ARA techniques may negatively impact application performance and user experience.

\begin{table}[]
\caption{Code Coverage Rates Under Different Difficulty Levels for the Original Apks}
    \centering
    \renewcommand{\arraystretch}{1.25}
    \resizebox{0.5\textwidth}{!}{ 
    \begin{tabular}{ccccc}
        \toprule
        \textbf{Metric (\%)} & \textbf{EASY} & \textbf{NORMAL} & \textbf{HARD} & \textbf{CHALLENGING} \\
        \midrule
        Mean       & \textcolor{red}{36.49} & 9.77 & 7.65 & \textcolor{orange}{5.58} \\
        Median     & \textcolor{red}{38.11} & 12.32 & 8.34 & \textcolor{orange}{5.65} \\
        1st Quartile & \textcolor{red}{12.58} & 8.12 & 5.51 & \textcolor{orange}{3.65} \\
        4th Quartile & \textcolor{red}{94.55} & 14.71 & 9.75 & \textcolor{orange}{7.29} \\
        \bottomrule
    \end{tabular}
    }
    \label{tab:Code coverage rates under different difficulty levels for the Original dataset}
    \vspace{-1.5em}
\end{table}

Table \ref{tab:Code coverage rates under different difficulty levels for the Original dataset} presents various code coverage metrics for the original dataset under different difficulty levels, including the mean, median, first quartile, and third quartile.  The experimental results on the original dataset more accurately reflect the impact of the number of ARA technology deployments on code coverage.

Combining the data from Figure \ref{fig:Code Coverage by Different ARA Quantity} and Table \ref{tab:Code coverage rates under different difficulty levels for the Original dataset}, it can be observed that the code coverage achieved by dynamic analysis tools decreases significantly as the number of deployed ARA techniques in the target APK increases. The highest values for all code coverage metrics appear in the EASY dataset, while the lowest are found in the CHALLENGING dataset. Additionally, as the difficulty level increases, each metric shows a consistent downward trend. Table \ref{tab:Code coverage rates under different difficulty levels for the Original dataset} further illustrates that with each increase in difficulty level, the coverage metrics decline markedly.

To evaluate whether application complexity—independent of ARA techniques—could have contributed to the observed differences, we also examined the average app size for each category. The results show a gradual increase across the four datasets: 8.35 MB (EASY), 8.65 MB (NORMAL), 8.97 MB (HARD), and 9.02 MB (CHALLENGING). While app size does tend to increase alongside ARA deployment level—possibly reflecting added complexity—the magnitude of increase is relatively minor. Notably, there is a clear gap in code coverage between the EASY and NORMAL categories, despite their similar average sizes. This suggests that app size alone is unlikely to explain the observed coverage degradation.

\vspace{0.2em}

Based on the data from the charts, we believe that deploying ARA technology at the NORMAL level or above can effectively enhance an application's resistance to dynamic analysis, thereby ensuring runtime security.  It should be noted, however, that deploying too many ARA techniques may introduce significant overhead, potentially degrading the user experience.  We will continue to investigate this issue in future work to determine the optimal number of ARA deployments, aiming to achieve the best balance between performance and security.

\vspace{0.2em}

Furthermore, by observing the performance of various dynamic analysis tools across different difficulty levels, we have reached the following conclusions. APIMonitor is almost ineffective on datasets of any level. Both AndroidSlicer and ESdroid demonstrate significant performance declines at the HARD level, which suggests that they are better suited for analyzing APKs at EASY and NORMAL levels. DroidCat shows a severe performance drop on the CHALLENGING dataset, making it more suitable for analyzing APKs below the CHALLENGING level. In contrast, T-Recs and DroidDissector exhibit consistently stable performance across APKs of all difficulty levels, with DroidDissector showing particularly outstanding and reliable results.

\vspace{0.2em}

These results highlight the varying resilience of existing dynamic analysis tools under different protection levels,  emphasizing the critical and practical need to carefully match tool capabilities with the complexity of the target application being analyzed.

\rcomment{Q2 Finding}{-We found that dynamic analysis tools were most severely impacted by VED technology, resulting in the lowest code coverage.   In contrast, tools were least affected by AH technology, allowing for relatively higher coverage.Additionally, the number of ARA techniques deployed significantly impacts the code coverage achievable during dynamic analysis.}

\begin{tcolorbox}[left=2mm, right=2mm, top=1mm, bottom=1mm]
\faLightbulbO \ \  {\textbf{Q3: How efficient are dynamic analysis tools when processing a large-scale dataset of applications protected by ARA techniques?}}
\end{tcolorbox}

In Question 3, we aim to assess the overall efficiency of dynamic analysis tools when processing a large-scale dataset of applications protected by ARA techniques. For example, in our empirical study, each tool was executed on a dataset comprising 1,615 APKs, which we consider representative of a large-scale dynamic analysis scenario. This stands in contrast to small-scale evaluations, where only a handful of selected samples are analyzed, and execution time or resource consumption may be less critical.

The efficiency of each tool is evaluated from two key aspects under large-scale analysis conditions.   The first is the time consumption during the preprocessing phase, which serves as a direct indicator of a tool’s runtime performance.   The second is the success rate of preprocessing and installation;   failure to produce a valid, executable output effectively invalidates the analysis effort.   Given the inherent time and resource intensiveness of dynamic analysis—particularly on a large-scale dataset—tool scalability and stability become essential factors influencing practical adoption and usability.

While it is acknowledged that the tools listed in Table~\ref{tab:Dynamic analysis tools} serve different primary purposes, some—such as APIMonitor, DroidCat, and DroidDissector—offer overlapping functionalities and are often employed in similar types of dynamic analysis workflows. Therefore, comparative insights regarding time efficiency and tool reliability remain valuable. For example, users can use this information to estimate expected preprocessing time, assess the tool's stability across benign and malicious apps, and determine whether the time cost is acceptable under constrained analysis scenarios. These insights can support more informed decisions regarding resource allocation and tool selection in practice.

Moreover, identifying cases where a tool demonstrates limitations—such as consistent preprocessing failures or substantial delays—can inform both practical adjustments (e.g., pre-filtering samples) and future tool improvements. Thus, the evaluation not only benchmarks performance but also highlights actionable insights for users and developers alike.

Table \ref{table:Average Installation Time and Success Rate of Tools} presents the time consumption of each tool during the preprocessing phase and whether they can output a APK file without errors during this phase.

\begin{table*}[h!]
\centering
\caption{Average Pretreatment Time and Success Rate of Tools}
\label{table:Average Installation Time and Success Rate of Tools}
\renewcommand{\arraystretch}{1.25}
\begin{threeparttable}
    \begin{tabular}{ccccccc}
    \toprule
    \multicolumn{1}{c}{\textbf{\multirowcell{2}[0pt][c]{Tool}}} & \textbf{Ben. Avg.} & \textbf{Ben. Success} & \textbf{Mal. Avg.} & \textbf{Mal. Success} & \textbf{Overall Avg.} & \textbf{Overall Success} \\
                       & \textbf{Time (s)}       & \textbf{Rate (\%)} & \textbf{Time (s)}  & \textbf{Rate (\%)}     & \textbf{Time (s)}     & \textbf{Rate (\%)} \\
    \midrule
    APIMonitor & \textcolor{orange}{3.26} & 86.1 & \textcolor{orange}{5.43} & \textcolor{orange}{72.3} & \textcolor{orange}{4.34} & 79.2 \\
    AndroidSlicer & 310.88 & 75.2 & 57.83 & 80.3 & 184.36 & 77.8 \\
    DroidCat & 48.37 & \textcolor{orange}{41.2} & 20.10 & 92.3 & 34.20 & \textcolor{orange}{66.8} \\
    T-Recs & 129.82 & 95.0 & 50.83 & 90.3 & 90.32 & 92.7 \\
    DroidDissector & 25.23 & \textcolor{red}{100.0} & 9.20 & \textcolor{red}{99.5} & 17.22 & \textcolor{red}{99.8} \\
    ESdroid & \textcolor{red}{327.16} & 75.5 & \textcolor{red}{58.56} & 80.3 & \textcolor{red}{192.85} & 77.9 \\
    \bottomrule
    \end{tabular}
\end{threeparttable}
\vspace{-1.5em}
\end{table*}

Among the tools, APIMonitor achieved the shortest time consumption on both benign and malicious datasets, with an average time of 4.34 seconds. In contrast, ESdroid had the highest time overhead, with an average time of 192.85 seconds on both types of datasets. Regarding preprocessing success rates, DroidCat had the lowest success rate on the benign dataset at only 41.2\%, but its success rate on the malicious dataset was 92.3\%. This discrepancy may be attributed to two possible explanations: one possibility is that DroidCat, being a dynamic behavior capture tool, is more suited for malicious APKs; another possibility is that DroidCat struggles with larger APK files. APIMonitor had the lowest preprocessing success rate on the malicious dataset at 72.3\%. On both types of datasets, DroidDissector achieved the highest preprocessing success rate, with an average success rate of 99.8\%. Specifically, DroidDissector achieved a 100\% preprocessing success rate on the benign dataset.

In Table \ref{table:Installation and Operate Success Rate}, the success rate indicates the probability that an APK can be properly installed and run smoothly on an emulator after being pre-processed by a dynamic analysis tool. Only APK files that are valid and usable for dynamic analysis are meaningful to researchers. Accordingly, this metric serves as a critical indicator of tool efficiency.

Among them, APIMonitor achieved the lowest success rate on both types of datasets, with an overall success rate of only 20.5\%.   Apart from APIMonitor, AndroidSlicer and ESdroid had the next lowest success rates, with success rates of only 56.8\% and 63.9\%, respectively.   This means that approximately half of the APKs cannot be correctly used for subsequent dynamic analysis, and the success rates of these two tools differ significantly compared to others.   Both AndroidSlicer and ESdroid had significantly lower success rates on benign datasets compared to malicious datasets.   Among the benign datasets, DroidDissector achieved the highest success rate at 87.6\%.   For the malicious datasets, T-Recs achieved the highest success rate, reaching an impressive 96.2\%.   Furthermore, T-Recs also had the highest overall success rate among all the tools, with an overall success rate of 91.1\%.

Considering the data presented in Table \ref{table:Average Installation Time and Success Rate of Tools} and Table \ref{table:Installation and Operate Success Rate}, T-Recs and DroidDissector demonstrate excellent overall efficiency.   For large-scale analysis datasets, researchers can prioritize using these two tools to enhance their research efficiency.

\rcomment{Q3 Finding}{-APIMonitor is unable to effectively handle newer versions of APK files, while the combined efficiency of T-Recs and DroidDissector makes them more suitable for large-scale datasets.}

\begin{table}[h!]
\centering
\caption{Installation and Operate Success Rate}

\begin{threeparttable}
\renewcommand{\arraystretch}{1.25}
    \resizebox{0.5\textwidth}{!}{%
    \begin{tabular}{cccc}
    \toprule
    
    \multicolumn{1}{c}{\textbf{\multirowcell{2}[0pt][c]{Tool}}} & \textbf{Ben. Success} & \textbf{Mal. Success} & \textbf{Overall Success}  \\
                       & \textbf{Rate (\%)}       & \textbf{Rate (\%)} & \textbf{Rate (\%)}   \\
    \midrule
    
    APIMonitor & \textcolor{orange}{25.1} & \textcolor{orange}{14.9} & \textcolor{orange}{20.5}  \\
    AndroidSlicer & 36.3 & 76.1 & 56.8  \\
    DroidCat & 77.2 & 89.1 & 85.4  \\
    T-Recs & 86.3 & \textcolor{red}{96.2} & \textcolor{red}{91.1}  \\
    DroidDissector & \textcolor{red}{87.6} & 90.7 & 89.1  \\
    ESdroid & 40.4 & 86.1 & 63.9  \\
    \bottomrule
    
    \end{tabular}
    }
\end{threeparttable}

\label{table:Installation and Operate Success Rate}
\vspace{-1.5em}
\end{table}

\section{DISCUSSION}
\label{sec:DISCUSSION}
{

Our research highlights the persistent and multifaceted challenges that developers of dynamic analysis tools face in effectively coping with ARA techniques widely deployed in modern Android applications. The evaluation results underscore a critical and urgent gap: none of the six tested tools—regardless of their intended analysis purpose—demonstrated sufficient capability to overcome ARA techniques, in either benign or malicious apps. This limitation holds even when applications are instrumented and analyzed under favorable conditions (Route 1), indicating a systemic shortcoming in current dynamic analysis infrastructures.

From a technical standpoint, different ARA categories vary significantly in effectiveness. Among the five ARA types we examined, VED exhibited the strongest resistance to dynamic analysis. In contrast, AH techniques had the weakest practical impact—some failed to activate even under controlled analysis conditions. This discrepancy is explainable by deployment context and ecosystem support: VED enjoys official support from Android, which ensures standardized implementation and stable behavior. This advantage is reflected in its real-world prevalence: 92.6\% of benign applications and 33.0\% of malicious applications in our dataset implemented VED techniques. AH lacks any platform-level support and involves higher implementation complexity, which likely limits its deployment and reliability. These observations suggest that in-the-wild defensive strength of AH may be substantially less than anticipated.

Another key finding relates to the quantity of deployed ARA techniques. Applications that implemented even a small number of ARA measures showed marked increases in analysis resistance. The most pronounced decline in code coverage occurred when comparing applications with zero ARA techniques to those with 1–5 techniques. While further increasing the number of techniques (beyond five) continued to suppress coverage, the marginal benefit diminished. This observation suggests that deploying a moderate number (1–5) of carefully selected ARA techniques strikes an effective balance between protection and runtime performance, thus serving as a practical guideline for developers balancing security and performance.

From the standpoint of benign developers, these findings highlight ARA techniques' dual impact. On one hand, ARA implementation significantly hinders code inspection, providing crucial protection for intellectual property and helping mitigate reverse engineering–based vulnerabilities. On the other hand, overuse of such techniques introduces non-trivial runtime overhead, which may impair application responsiveness and user experience. Developers should therefore adopt a targeted deployment strategy—prioritizing highly effective, low-overhead techniques such as VED, and carefully evaluating trade-offs introduced by more complex mechanisms like AH. Strategic selection is essential to maintaining usability while enhancing security.

For malware analysts and forensic engineers, the implications are equally important. Our findings show that existing dynamic analysis tools fail to automatically bypass ARA protections, regardless of application type or tool design. In practice, this necessitates manual intervention in many cases—such as altering the execution environment, disabling protection routines, or manually patching out obfuscation logic. These tasks demand considerable expertise and effort, thereby increasing the barrier to scalable analysis. Even the best-performing tool in our evaluation, DroidDissector, while demonstrating relatively better code coverage and runtime performance, still struggled to overcome ARA techniques when multiple layers were deployed.

Although our evaluation primarily used code coverage as a proxy for dynamic analysis effectiveness, which does not fully represent a tool’s total functionality (e.g., behavior monitoring, network tracing, or anomaly detection), it remains a critical baseline indicator. Given the diverse purposes of the tested tools, achieving uniform evaluation remains difficult. Nonetheless, their collective failure to respond effectively to ARA techniques exposes a common weakness that limits the practical utility of dynamic analysis in modern software environments.

In light of these observations, it is imperative that future development of dynamic analysis tools emphasize robust, modular support for ARA detection and mitigation. The limitations exposed in our evaluation indicate that traditional models of behavior capture—typically built around system call tracing, API monitoring, and surface-level instrumentation—are insufficient in environments where ARA techniques have become both prevalent and increasingly sophisticated. Instead, dynamic analysis tools must evolve into flexible, extensible frameworks that can adapt to diverse protection strategies, actively detect signs of anti-analysis behavior, and respond with effective countermeasures.

Rather than relying solely on runtime observation, future tools should adopt hybrid analysis paradigms, integrating static and dynamic techniques in a coordinated manner. Static analysis can play a key role in identifying potentially protected regions, opaque predicates, or environmental dependency checks that may hinder execution. When combined with dynamic execution, this hybrid approach enables more targeted and efficient exploration of program logic. For instance, static control-flow graph (CFG) extraction can help isolate code blocks likely to be guarded by ARA techniques, guiding the dynamic engine to focus inputs or emulate specific runtime conditions. This not only improves code coverage but also reduces the time required for comprehensive behavior extraction.

Another promising direction lies in the integration of lightweight machine learning (ML) models. These models can be trained to recognize patterns of code obfuscation, behavioral anomalies, or unusual branching behavior indicative of ARA routines. During runtime, ML-assisted detection can trigger adaptive behaviors in the analysis tool—such as toggling instrumentation strategies, injecting crafted environment variables, or applying symbolic execution selectively. These capabilities allow the tool to move beyond reactive tracing and toward proactive analysis, anticipating and preempting ARA-based disruptions.

Equally essential is enhancing execution environments.   Configurable sandboxes must be designed to simulate a wide variety of device states, system contexts, and user behaviors, since many ARA techniques rely on environment checks to detect emulation or instrumentation. By providing fine-grained control over runtime parameters—including OS properties, sensor states, installed apps, or even execution timing—future tools can more effectively bypass or neutralize ARA checks. Additionally, advanced sandboxes should support real-time memory inspection, API call interception, and function hooking, allowing analysts to surgically intervene in execution when necessary.

To ensure that research progress is sustainable and measurable, the field must also prioritize the establishment of standardized evaluation benchmarks and annotated datasets. These resources should include applications embedded with diverse and well-documented ARA techniques, enabling comparative evaluation of tool effectiveness. Without such shared infrastructure, it is difficult to gauge improvements, identify persistent blind spots, or promote reproducible research across the community.

Finally, dynamic analysis advancement must coordinate closely with reverse engineering, automated deobfuscation, and adversarial modeling. These subfields offer complementary techniques—binary diffing, symbolic reasoning, fuzzing—to enhance ARA-aware analysis. Future tools need analytical power and adaptive intelligence, learning from failures and evolving with targeted defenses.

In a software ecosystem where ARA techniques are no longer exceptional but expected, the survival and utility of dynamic analysis depend on its ability to keep pace. By embracing hybrid architectures, intelligent automation, environment simulation, and collaborative benchmarking, the next generation of dynamic analysis tools can re-establish their relevance as essential instruments in the ongoing battle between software protection and security analysis.

}

\section{THREATS TO VALIDITY}
\label{sec:THREATS_TO_VALIDITY}

External validity threats concern the generalizability of our experimental results. The findings of this study are based on application samples obtained from AndroZoo. However, due to the limited number of analyzed applications and their market distribution, this research may not fully represent the entire Android application ecosystem. To mitigate this, we attempted to acquire and analyze as many applications as possible to enhance the universality of the results. Large-scale dynamic analysis is highly time-consuming; in our work, we performed multiple analyses on each APK file, with each analysis taking approximately 10 minutes. Despite these constraints, our dataset includes 993 benign and 991 malicious applications, and the complete analysis process took three months. Additionally, to reduce potential dataset contamination caused by outdated vt\_detection values in AndroZoo, we re-scanned all samples using VirusTotal to ensure the reliability of the dataset.

Internal validity threats relate to the reliability and effectiveness of the measurement tools used. In this study, we measured code coverage using ACVTool. Given the challenges of measuring fine-grained coverage in closed-source applications, the coverage results obtained may carry inherent biases, which are difficult to eliminate. Nevertheless, we believe these limitations do not hinder our ability to derive meaningful qualitative conclusions.

Furthermore, in our current implementation, ARAP computes the presence of different types of ARA techniques per APK, but does not count the number of times each technique is applied across different locations in the code. Repeated application of the same ARA technique across multiple code locations may increase analysis complexity. However, due to ARAP's current limitations, reliably detecting multiple instances of the same ARA technique remains a challenge. We consider this a limitation and plan to enhance the granularity of ARAP's detection in future work to better address this issue.

\section{Related Work}
\label{sec:RelatedWork}

To orient the reader, we organize prior work into four complementary strands that move from broad characterization to concrete countermeasures: 
(1) \emph{surveys and taxonomies of Android security tooling}, 
(2) \emph{empirical comparisons of analysis tools}, 
(3) \emph{studies of resilience to code transformations (e.g., obfuscation/anti-analysis)}, and 
(4) \emph{runtime approaches that attempt to counter ARA techniques}. 
Across these strands, most existing work either focuses on static analysis or does not evaluate contemporary \emph{dynamic} analysis tools under real-world anti–runtime-analysis (ARA) defenses. 
Our study fills this gap by providing a large-scale, tool-centric empirical evaluation of dynamic analysis tools confronted with diverse ARA categories, quantifying runtime visibility (coverage) and practical efficiency at scale.

\medskip
\noindent\textbf{(1) Summarizing and categorizing tool features.}
Several studies systematically catalog and compare features of Android security tools. 
\citet{Daoudi2022AssessingTO} evaluate state-of-the-art malware detection methods and discuss strategies for combining them to improve detection rates. 
\citet{Heid2021AutomatedDA} summarize characteristics of vulnerability detection tools with an emphasis on usability, and 
\citet{Baheux2023DroidSecTesterTC} analyze a large body of research from top venues to assess tool usability.

These studies offer a crucial foundational overview of the tooling landscape, aiding researchers and practitioners in understanding available options. However, their primary focus remains on theoretical feature comparison and high-level usability, and they do not include empirical evaluations of how these tools perform under real-world constraints.Specifically, they lack experimental data on a critical practical challenge: the tools' resilience against widespread ARA techniques. Our work seeks to build upon this foundational research by providing a comprehensive empirical assessment that quantifies tool effectiveness and efficiency when confronted with such protections.

\medskip
\noindent\textbf{(2) Measuring tool performance through experiments.}
A second line of work performs controlled, head-to-head comparisons of analysis tools, most often in \emph{static} settings. 
Zhang et al.~\citep{DBLP:journals/tse/ZhangWQR22} provide a comprehensive, controlled, and independent comparison of three prominent static analysis tools for information-flow (pollution): FlowDroid~\citep{arzt2014flowdroid}, Amandroid~\citep{wei2018amandroid}, and DroidSafe~\citep{gordon2015information}. 
Ranganath et al.~\citep{DBLP:journals/ese/RanganathM20} evaluate Android vulnerability detection tools using the Ghera benchmark (considering 64 tools, assessing 14). 
Bonett et al.~\citep{DBLP:conf/uss/BonettKMNP18} systematically evaluate static tools to discover, document, and fix defects, showing defect propagation across derivatives. 
Pauck et al.~\citep{DBLP:conf/sigsoft/PauckBW18} introduce BareDroid/ReproDroid to enable accurate comparisons among six information-flow tools (Amandroid~\citep{wei2018amandroid}, DIALDroid~\citep{dialdroid}, DidFail~\citep{didfail}, DroidSafe~\citep{gordon2015information}, FlowDroid~\citep{arzt2014flowdroid}, IccTA~\citep{li2015iccta}). 
Aloraini et al.~\citep{DBLP:conf/icsm/AlorainiN17} study six static tools for buffer-overflow detection and report limited effectiveness for advanced FOSS tools. 

These studies have established a strong methodological foundation for the empirical assessment of analysis tools. However, their focus has been predominantly confined to the domain of static analysis. Furthermore, these evaluations are conducted from a traditional perspective, assessing tools under ideal conditions rather than against the sophisticated and widespread ARA techniques deployed in modern applications. Consequently, it remains unclear how well these tools, and particularly dynamic analysis tools which are renowned for their accuracy but are also highly susceptible to runtime environments, would perform when such protections are present. Our work directly addresses this gap by conducting the first large-scale empirical evaluation of dynamic analysis tools specifically under the challenge of ARA techniques.

\medskip
\noindent\textbf{(3) Evaluating the ability of tools to counter code transformations.}
A third strand investigates how obfuscation or related transformations affect analysis or detection quality. 
Hammad et al.~\citep{DBLP:conf/icse/HammadGM18} conduct a large-scale empirical study showing that code obfuscation significantly impacts Android anti-malware products. 
Soi et al.~\citep{DBLP:journals/corr/abs-2312-17356} propose a visibility metric to assess the difficulty of detecting non-operational code (e.g., NOPs) and discuss the stealthiness of evasive applications. 
Nawaz et al.~\citep{DBLP:journals/peerj-cs/NawazAL22} evaluate anti-malware tools under complex hybrid obfuscation and observe substantial effects on detection rates. 

These studies have established rigorous methodologies for evaluating tool resilience against code transformations and yielded valuable insights into obfuscation's impact on static analysis. However, their scope remains limited to static analysis and malware detection, with the broader spectrum of ARA techniques that threaten dynamic analysis still largely unexplored. Our work extends this important research direction by conducting a comprehensive evaluation of dynamic analysis tools against a wide range of ARA techniques, moving beyond obfuscation to provide a holistic view of their defensive capabilities.

\medskip
\noindent\textbf{(4) Addressing new attempts to counter ARA techniques.}
A final strand proposes runtime mechanisms to tame or bypass ARA defenses. 
Wang et al.~\citep{Wang2017DroidAntiRMTC} present Droid-AntiRM, which detects anti-analysis checks and manipulates bytecode to force specific runtime behaviors. 
Tang et al.~\citep{Tang2018DualForceUW} introduce Dual-Force, simultaneously forcing Java and JavaScript (WebView) execution paths to expose hidden payloads without manual input. 
Wang et al.~\citep{Wang2019AutomatedHA} propose DirectDroid, which enhances fuzzing via on-demand enforcement to bypass checks and redirect execution to target locations. 

These studies provide valuable conceptual frameworks and pave the way for future research. However, a significant barrier to practical adoption and independent validation exists: the tools themselves are not publicly available. The lack of open-source implementations or readily deployable systems makes it difficult to benchmark these novel approaches against existing tools or to integrate their insights into practical workflows. Our future work aims to reconstruct these tools to facilitate continued research progress.

In contrast to prior surveys, static-focused benchmarks, and obfuscation-centric or non-public counter-ARA proposals, our work provides the first large-scale,  tool-centric assessment of contemporary dynamic analysis tools confronted with diverse ARA categories, measuring both runtime coverage (visibility) and practical efficiency across thousands of runs.

\section{CONCLUSION}
\label{sec:CONCLUSION}

In this study, we conducted an initial evaluation of existing Android dynamic analysis tools in handling ARA techniques, using code coverage as the primary assessment metric. This research addresses a notable gap in the literature on ARA techniques and dynamic analysis tools, offering valuable insights to guide the future development of security protection technologies. Our findings reveal that while current dynamic analysis tools fall short in addressing ARA techniques, they exhibit promising potential for enhancement. Future work may explore approaches to improve the adaptability of dynamic analysis tools to ARA techniques and enhance their practical effectiveness. Overall, this study deepens our understanding of Android application security and anti-analysis mechanisms, contributing meaningfully to both academic research and real-world security practices.

\section*{Acknowledgements}
The authors would like to thank the anonymous reviewers for their valuable comments and helpful suggestions. We acknowledge the support received from the National Natural Science Foundation of China (No. 62472234, No. 62372245), and the Natural Science Foundation of Xinjiang Uygur Autonomous Region, China under the grant number of 2024D01A55.

\bibliographystyle{model2-names}

\bibliography{References}




\end{document}